\documentclass[reprint,superscriptaddress,preprintnumbers,nofootinbib,nobibnotes,amsmath,amssymb,aps,prl,floatfix,noeprint]{revtex4-2}

\usepackage{amsmath,amsbsy,esint,mathrsfs,commath,amsfonts,mathtools,amsthm}
\usepackage[usenames,dvipsnames]{xcolor}
\usepackage{braket}
\usepackage{tensor}
\usepackage{physics}
\usepackage{graphicx}
\usepackage{xparse}
\usepackage{hyperref}
\usepackage[caption=false]{subfig}

\usepackage[normalem]{ulem}

\usepackage{overpic}

\hypersetup{
    colorlinks=true,
    linkcolor=Blue,
    filecolor=magenta,      
    urlcolor=Blue,
    citecolor=Blue,
    }

\newcommand{\Ea}{E^{\alpha}}
\newcommand{\Eb}{E^{\beta}}
\newcommand{\Fa}{F^{\alpha}}

\def\constr{L}

\def\bi{\beta}
\def\constA{\Sigma}
\newcommand{\Fb}{F^{\beta}}
\def\SU{\mbox{SU}}

\def\U{\mbox{U}}

\newcommand{\N}{\mathcal{N}}

\def\ham{H_0}
\def\amin{a_\text{gap}}
\def\aamin{A_\text{gap}}
\def\bb{\textsc{bb}}

\begin{document}

\title{Global variables and dynamics of emergent cosmology in loop gravity}

 \date{\today}
  
 \author{I\~naki Garay} 
 \email{inaki.garay@ehu.eus}
 \affiliation{Department of Physics and EHU Quantum Center,
   University of the Basque Country EHU, 48940 Leioa, Basque Country, Spain}  
   
\author{Luis J. Garay} 
\email{luisj.garay@ucm.es}
	\affiliation{Departamento de F\'{\i}sica Te\'orica \& IPARCOS, Universidad Complutense de Madrid, 28040 Madrid, Spain}
	
\author{Diego H. Gugliotta}
    \email{diego.henriquez@uam.es}
    \affiliation{Departamento de F\'{\i}sica Te\'orica de la Materia Condensada and Condensed Matter Physics Center (IFIMAC), Universidad Aut\'onoma de Madrid, 28049 Madrid, Spain}
    \affiliation{Instituto de Qu\'imica F\'isica Blas Cabrera (IQF), CSIC, 28006 Madrid, Spain}
    
 \author{Sergio Rodr\'iguez-Gonz\'alez}
 \email{sergio.rodriguez@ehu.eus}
 \affiliation{Department of Physics and EHU Quantum Center,
   University of the Basque Country EHU, 48940 Leioa, Basque Country, Spain} 
   
 \author{Ra\"ul Vera} 
 \email{raul.vera@ehu.eus}
 \affiliation{Department of Physics and EHU Quantum Center,
   University of the Basque Country EHU, 48940 Leioa, Basque Country, Spain}  
  
\begin{abstract}
Deriving emergent cosmology from the discrete geometry of Loop Quantum Gravity remains an open challenge. We solve the two-vertex model exactly, including general non-symmetric configurations, by introducing macroscopic global variables that form a closed Poisson subalgebra. This yields an intrinsic minimal area gap from the 
Casimir invariants of the subalgebra. In the semiclassical limit, a generalized Friedmann equation emerges incorporating graph anisotropies and quantum geometry corrections, which lead to a Big Bounce. This global formulation opens a coarse-graining route toward statistical descriptions of quantum spacetime.
 \end{abstract}
 
\maketitle

\paragraph{Introduction.---}
One of the most remarkable expectations in quantum gravity is the emergence of a fundamental discreteness of geometry at the Planck scale \cite{Garay:1994en,Hossenfelder:2012jw}. In Loop Quantum Gravity (LQG), this feature is encoded in the discrete spectra of geometric operators, which in particular imply the existence of a non-vanishing area gap. Although the kinematical framework of the theory is now well established, deriving and solving its quantum dynamics, as well as understanding how the correct semiclassical limit emerges, remain among its central open challenges \cite{Thiemann,Rovelli}. 
In this context, simple models provide a valuable setting in which these challenges can be addressed in a controlled way while retaining essential features of the full theory. They also offer a natural framework to explore the relation between the full theory and symmetry-reduced models such as Loop Quantum Cosmology, whose precise connection with LQG remains an open question \cite{Ashtekar:2021kfp}.

In the canonical formulation of LQG, the kinematical Hilbert space is spanned by spin-network states defined on graphs, which provide a simultaneous eigenbasis for the geometric operators of area and volume \cite{Rovelli:1994ge,*RovelliErratum,AreaOperator}. The spectrum of the area operator is determined by the spins associated with the links of the graph, while the volume operator acts non-trivially on the intertwiners assigned to its vertices. At the classical level, fixing a graph yields a finite-dimensional phase space naturally described in terms of twisted geometries \cite{TwistedGeometries,TwistorsTG}. 
Within this framework, the fundamental variables are associated with the links and encode both intrinsic and extrinsic geometric data, whereas each vertex admits the interpretation of a polyhedron whose faces are dual to the adjacent links. Twisted geometries generalize Regge geometries \cite{Tambornino3,TwistedGeometries,TwistorsTG} by allowing discontinuities in the shapes of shared faces of neighboring polyhedra, only preserving the matching of their areas. 

A specially fruitful simple model is the truncation to a fixed graph given by the so called two-vertex model, that consists of
two vertices connected by $N$ oriented links
\cite{Rovelli2vertex,returnSpinor,2vertexN,Livine:2011up,Aranguren:2022nzn,Cendal:2024uzu,Garay:2025bqk,Garay:2025cis,Assanioussi:2026cee,Garay:2026xzn}.
This model has established a robust relation between LQG and cosmology \cite{Livine:2011up,returnSpinor,2vertexN,Aranguren:2022nzn,Garay:2025cis,Cendal:2024uzu}: 
 the symmetry reduced sector given by imposing suitable  $\U(N)$ constraints
yielded a direct correspondence with the Friedmann–Lemaître–Robertson–Walker (FLRW) geometry and its dynamics \cite{Cendal:2024uzu,Garay:2025cis}, while deviations from this sector provide a natural framework to study anisotropies and inhomogeneities \cite{Aranguren:2022nzn,Garay:2025bqk}. However, the full analytic study for the model in the general $\U(N)$ non-symmetric sector has not been accomplished yet, and only  numerical results have been obtained \cite{Aranguren:2022nzn}.

In this Letter we analytically solve the global dynamics of the two-vertex model of LQG in full generality, by introducing a set of global variables with direct geometric interpretation. 
An area gap for the graph emerges naturally from this construction. These global variables also yield conserved quantities directly tied to the $\U(N)$ constraints, clarifying the role of symmetry and its breaking in the model. They simplify the equations of motion considerably. Most remarkably,
we find that the equation for the scale factor (the square root of the total area) takes, in the general case,  the same form of the generalized Friedmann equation found in \cite{Cendal:2024uzu} for the $\U(N)$-reduced sector, plus an additional term proportional to one of the conserved quantities encoding the anisotropies of the graph.
This conserved quantity allows us to characterize the anisotropic direction of one of the reduced sectors treated in \cite{Garay:2025bqk} in terms of an intrinsic three-dimensional vector. 

In short, this new formulation provides a complete analytical characterization of the model from which cosmological dynamics naturally emerges. It also paves the way for systematic coarse-graining schemes in which the two-vertex model serves as an elementary building block for exploring the semiclassical limit of increasingly refined graphs \cite{Assanioussi:2026cee}.

\paragraph{Two-vertex model.---}
We use the spinorial formalism \cite{returnSpinor,Tambornino1,Tambornino2,
Tambornino3} to parametrize the phase space of the
two-vertex model:
each link $i=1,\dots,N$ of the graph is endowed with a pair of spinors 
$z_i^{\alpha A}$ and $z_i^{\beta A}$
attached to vertices $\alpha,\,\beta$, respectively, where $z^A=z^0, z^1\in \mathbb{C}$,
satisfying the Poisson algebra
\begin{equation}
    \big\{z^{\nu A},\bar{z}^{\nu B}\big\}=-i\delta^{AB},
  \label{eq_zPoissonBrackets}
\end{equation}
and zero the rest, where $\nu=\alpha,\beta$.

These spinors define a set of $\SU(2)$-invariant observables at each vertex, given by
\begin{equation}
E^\nu_{ij} :=\bar{z}^\nu{}^0_i z^\nu{}_j^0+\bar{z}^\nu{}^1_i z^\nu{}_j^1,
\qquad F^\nu_{ij} := z^\nu{}_i^0 z^\nu{}_j^1-z^\nu{}_i^1 z^\nu{}_j^0,
\label{eq:EFs}
\end{equation}
which, in turn, satisfy a closed Poisson algebra,  and provide a complete description of the gauge-invariant degrees of freedom at each node \cite{Girelli:2005ii,2vertexN,returnSpinor,Girelli:2017dbk}.
The diagonal components $E^\nu_{ii}$ are related to the areas of the faces, 
while the off-diagonal components encode information about their relative orientation. Indeed, the closure constraint which ensures the existence of a convex polyhedron at each vertex can be written as \cite{returnSpinor}
\begin{equation}
\sum_i \bar{z}^\nu{}_i^A z^\nu{}_i^B =\frac{1}{2}\sum_{i}E^\nu_{ii}\delta^{AB}=:A^\nu\, \delta^{AB},
\label{closure_constraint}
\end{equation}
where $A^\nu$ is the total area of the polyhedron corresponding to the
vertex $\nu$,
which is non-negative by construction. 
 In addition, one imposes matching conditions 
along each link $E^\alpha_{ii}=E^\beta_{ii}$ to ensure that the areas associated with the two endpoints is the same. Of course
this implies, in particular, that the total
areas of the two polyhedra coincide, and we write $A:=A^\alpha=A^\beta$.
The way two adjacent faces are joined
is encoded in variables called twist angles, related with the presence of extrinsic curvature degrees of freedom
\cite{TwistedGeometries,TwistorsTG,returnSpinor,2vertexN,Livine:2011up,Aranguren:2022nzn,Cendal:2024uzu,Garay:2025bqk,Garay:2025cis}.

This spinorial framework is a powerful tool to  study the classical dynamics of simple models of LQG. In particular, it allows the definition of $\SU(2)$-invariant interactions between 
vertices in terms of $E^\nu_{ij}$ and $F^\nu_{ij}$, which serve as the starting point for the dynamical analysis.

Non-trivial dynamics on the two-vertex model has been implemented using the Hamiltonian \cite{Cendal:2024uzu,Garay:2025bqk,Garay:2025cis}
\begin{equation}
    \ham = \sum_{i,j } \left(\lambda \Ea_{ij}\Eb_{ij} + \gamma\Re( \Fa_{ij} \Fb_{ij})\right), 
      \label{hamiltoniano}
\end{equation}
where $\lambda  \in \mathbb{R}$, $\gamma\in\mathbb{R}^+$ are coupling constants (we can choose $\gamma\in\mathbb{R}^+$ without loss of generality because its complex phase can be absorbed by shifting the twist angles between the faces \cite{Garay:2025bqk}).
This Hamiltonian is the leading nontrivial term for a compatible LQG dynamics of the two-vertex truncation. Indeed, it 
commutes under Poisson brackets with the closure and matching constraints, and includes, up to regularization ambiguities, the contributions from all holonomies on the graph \cite{Rovelli2vertex,Aranguren:2022nzn}.
Note that, with  an appropriate choice of the coupling constants, the Hamiltonian $\ham$
reduces to a form closely related to that proposed in \cite{Rovelli2vertex} for the two-vertex model, whose regularization is based on the original LQG Hamiltonian operator construction of \cite{RovelliSmolinHamiltonian}. 

 \paragraph{Global variables.---}
\label{sec:global}
It is remarkable
that despite the sheer number of
variables required to describe the
whole collection of spinors in the model,
its global behavior is encoded in
the area $A$ together with the real variables $e$, $r$, and $\phi$ defined by
\begin{equation}
    e := \sum_{i,j}E_{ij}^{\alpha}E_{ij}^{\beta},\qquad  re^{i\phi} :=\sum_{i,j}F_{ij}^{\alpha}F_{ij}^{\beta}.
\end{equation} 
These variables are global in the sense that they do not carry information of specific links or nodes,
and they capture the main features of the geometry of the model, in particular, the total area $A$
and information on the anisotropy of the graph.
They satisfy, cf. \eqref{eq_zPoissonBrackets}-\eqref{closure_constraint},
\begin{equation}
        \left\{e,\phi\right\} = 4A,\quad \left\{r,\phi\right\} = 4A\frac{e}{r},\quad \left\{A,\phi\right\} = 1,\label{eq:globalPoisson}
\end{equation}
and zero the rest. Observe that this allows to interpret $\phi$ as a global twist angle, since it is conjugate to the total area \cite{TwistedGeometries}. 
A crucial fact is that Eqs. \eqref{eq:globalPoisson} 
define a closed Poisson subalgebra of the whole two-vertex system  with two Casimir operators. 
Let us see this in more detail. 

In terms of the generators of the $\U(N)$ symmetry
$\mathcal{E}_{ij} := E_{ij}^{\alpha} - E_{ji}^{\beta}$,
whose vanishing defines the $\U(N)$ constraints \cite{2vertexN}, we consider the
non-negative quantities
\begin{equation}
\constA \!= \frac{1}{4}\sum_{i,j}|\mathcal{E}_{ij}|^{2}, \label{constA}
  \qquad\constr\! = |\vec{L}|,
\end{equation}
where the $\mathbb{R}^3$-vector
\begin{equation}
\vec{L}:=\frac{1}{2}\sum_{ij}\mathcal{E}_{ij}\bar{z}^{\alpha A}_j\vec{\sigma}_{AB}z^{\alpha B}_i,
\label{eq:L}
\end{equation}
with $\vec{\sigma}$ being Pauli matrices,
provides a privileged direction intrinsically defined by all the spinors.
An analogous vector with the same modulus can be defined for the vertex $\beta$. The relation between the two vectors is  fixed by the graph itself.

Let us note that $\mathcal{E}_{ij}$ satisfy the $\U(N)$ algebra by construction,
and $\constA$ is
the quadratic Gel'fand invariant of $\U(N)$ (see e.g. \cite{Freidel_Livine_2010}).
A key feature of the two-vertex model is that,
quite unexpectedly, this invariant of $\U(N)$ is also
a Casimir of the subalgebra of the global variables.
Indeed, using the identity \mbox{${F}_{ij}\bar{F}_{lk} = E_{li}E_{kj} - E_{lj}E_{ki}$}
on each vertex, it is not difficult to see that $\constA$, as well as
$\constr$, can be written solely in terms of the  Poisson subalgebra elements above as
\begin{equation}
 \constA = A^{2}-\frac{e}{2}, \qquad  \constr = \frac{1}{2}\sqrt{e^{2}-r^{2}},
 \label{eq:constants}
\end{equation}
which are well defined by construction and therefore \mbox{$0\leq r\leq e\leq 2A^2$}. It is now straightforward to check that $\constr$ and $\constA$ are two Casimirs of the Poisson subalgebra \eqref{eq:globalPoisson}.
The relations above endow the two Casimirs with a clear physical interpretation in terms of an area gap for the graph. Indeed, Eq. \eqref{eq:constants} implies
\begin{equation}
A^2\geq\constA+\constr =: \aamin^{2}\,. \label{eq_minimalarea}
\end{equation} 
Observe that $\constr$ vanishes when $\constA$ vanishes, and the latter happens if  and only if the $\U(N)$ constraints $\mathcal{E}_{ij} = 0$ are imposed.
Therefore $\constA=0$ completely characterizes the $\U(N)$ sector, which is entirely equivalent to setting $\aamin = 0$.  

There exist other symmetry-reduced sectors \cite{Garay:2025bqk}.
The simplest example is a slight deviation from the $\U(N)$ sector of the two-vertex model with four links,
by allowing the generators $\mathcal{E}_{12}$, $\mathcal{E}_{13}$, and $\mathcal{E}_{14}$ 
to be different from zero. In that case, the unitary transformation relating the spinors at both ends of the link $1$ is different from the unitary transformation relating those at the ends of the links $2$, $3$ and $4$, so that the link $1$ constitutes a privileged direction in the graph.
It can be shown that in this case $\constA = \constr$. Then $\vec{L}$ is given by
\begin{equation}
\vec{L} = \frac{\aamin^{2}}{2}\hat{X}^{\alpha}_{1},
\end{equation}
where $\hat{X}^{\alpha}_{1}$ is the unit vector normal to the face $1$ of the polyhedron \cite{TwistedGeometries}. That is,
the vector $\vec{L}$ is orthogonal to the face of the polyhedron located at the node $\alpha$ associated with the privileged link (number 1). This suggests that the vector $\vec{L}$ may be interpreted as an indicator of anisotropies in the graph. Observe that
this is consistent with the fact that in the homogeneous and isotropic sector, $\mathcal{E}_{ij}=0$, and thus
$\vec{L} = 0$ by construction.

In terms of the global variables, the Hamiltonian \eqref{hamiltoniano} of the system takes the form
\begin{equation}
    \ham = \lambda e + \gamma\, r\cos{\phi}.\label{eq:ham}
\end{equation}
Furthermore, we may write $e$ and $r$ in terms of the two Casimirs $\constA$ and $\constr$ and the total area $A$,
reducing the phase space to a cylinder 
coordinated by $A$ and $\phi$. Then  the Hamiltonian \eqref{eq:ham} reads
\begin{equation}
    \ham = 2\lambda (A^{2}-\constA ) + 2\gamma\sqrt{(A^{2}-\constA)^{2} - \constr^2} \cos{\phi}.\label{Hamiltoniano_A_Phi}
\end{equation}
Moreover, since we are dealing with a conservative Hamiltonian system, the implicit form of the trajectories is given by Eq. \eqref{Hamiltoniano_A_Phi}.
Let us stress that this equation holds independently of the number of links of the graph and of the precise spinorial degrees of freedom that underlie the global picture. Most remarkably, Eq. \eqref{Hamiltoniano_A_Phi} allows the analytical identification of the regimes where area divergencies can arise in terms of the coupling constants, since it  implies 
\begin{equation}
    \frac{\ham^{2} + 4\constr^2\gamma^{2}}{4(A^{2}-\constA)^{2}} - \frac{ \lambda \ham}{(A^{2}-\constA)} \leq \gamma^{2}-\lambda^2.\label{inecuacion_divergent}
\end{equation}
Therefore, to be able to approach arbitrarily large values of the area $A$, it is necessary that the right hand side be non-negative, i.e. that $\gamma \geq |\lambda|$, which defines what has been called the divergent regime in the literature,
in contrast with the oscillatory regime for $\gamma<|\lambda|$,
all found previously only numerically \cite{Aranguren:2022nzn}.

Let us recall that the minimum bound for the area \eqref{eq_minimalarea} is independent of the value of the coupling constants $\lambda, \gamma$. As we discuss in the following, the formulation in terms of the global variables extends the cosmological correspondence of the two-vertex model beyond the symmetry-reduced $\text{U}(N)$ sector. This bound admits a sensible interpretation in terms of the Big-Bounce found in quantum cosmology \cite{Ashtekar:2011ni,Bojowald:2008zzb,Agullo:2016tjh}, as we make precise below.

Another important direct consequence of the bound \eqref{inecuacion_divergent} is that
the Hamiltonian can be taken to be a constraint (to ensure time-reparameterization invariance), $\ham=0$,
only  in the divergent regime $\gamma \geq |\lambda|$.

\paragraph{Generalized Friedmann equation.---}
\label{sec:Friedmann}

Motivated by the interpretation of $A$ as an area, and in order
to explicitly connect with cosmology, we perform the canonical transformation \cite{Cendal:2024uzu,Garay:2025cis}
\begin{equation}
    a = \sqrt{\bi A},\qquad \pi_a= 2\sqrt{A/\bi}\phi,
\end{equation}
where $\bi\in \mathbb{R}^+$ corresponds to the Barbero-Immirzi parameter,
characteristic of LQG \cite{Thiemann,Rovelli}.
Moreover we can also consider gauge invariant (under the action of the matching and closure constraints) modifications to the Hamiltonian
which entail a modification of the lapse function and the
addition of a term that allows for the inclusion of matter in the model \cite{Livine:2011up,Cendal:2024uzu,Garay:2025cis}.
In this scenario we introduce the Hamiltonian
\begin{equation}
    H=\frac{\N a}{(a^4-\beta^2 \Sigma)} \ham + \N a^3 f,\label{eq:H}
\end{equation}
for some arbitrary function $f(a)$.
Let us stress that this last term could have been introduced naturally
from the beginning in terms of the full set of $\SU(2)$-invariant
objects, as shown in \cite{Garay:2025cis}, and it is straightforward to check that
the conservation of $\constA$ and $\constr$ is unaffected. We take the positive quantity (see \eqref{eq_minimalarea})
$\N a/(a^4-\beta^2 \Sigma)$ as a Lagrange multiplier, chosen at will to fix a time parametrization, 
and now $H$ is a constraint. Note that the chosen lapse function  depends on $\constA$. Therefore, this Casimir and hence the $\U(N)$ symmetry naturally encode the foliation (or the observer's clock) used.

Computing $\dot a=\{a,H\}$ directly yields
\begin{equation}
    \frac{\dot a}{a}=
    - \N 
    \frac{\gamma}{\bi^2}
    \sqrt{1-\bi^4\constr^2\mu^2(a)}\sin\phi,
    \label{eq:dot_a}
\end{equation}
where
\begin{equation}
    \mu (a):= \frac{1}{a^{4}-\bi^{2}\constA}=\frac{1}{a^{4}-\amin^{4}+\bi^2\constr},
\end{equation}
and $\amin:=\sqrt{\bi \aamin}$. 
It is convenient to introduce
\begin{equation}
\kappa:=-\frac{2}{\beta^2}\gamma(\gamma+\lambda),
\label{eq:kappa}
\end{equation}
which 
is  non-positive 
if $f(a)=0$, because $\gamma\geq|\lambda|$ is required for the Hamiltonian to be a constraint. In general, 
however, the divergent and oscillatory regimes depend on $f$,
so $\kappa$ may also be positive for suitable values of the couplings $\gamma$, $\lambda$.
Using the time-reparameterization constraint \mbox{$H=0$}, and introducing the dimensionless quantity
\begin{equation}
\Psi (a):= \left(-\frac{\kappa}{a^{2}}+\gamma f(a)\right)\frac{a^{2}}{4\gamma^2},
\label{eq:Psi-def}
\end{equation}
 Eq. \eqref{eq:dot_a} acquires the form
\begin{equation}
\left(\frac{\dot a}{\N a}\right)^{2}=\frac{4\gamma^{2}}{a^{2}}\left(\Psi(1-\bi^2\Psi)-\frac{\bi^2\constr^2\mu^{2}}{4}\right).
\label{eq:master}
\end{equation}
This equation for $a(t)$ corresponds to a generalization of the Friedmann equation in FLRW, extending to the full two-vertex model the equation found in \cite{Cendal:2024uzu,Garay:2025cis}
for the $\U(N)$-sector by only adding the term proportional to $\constr^2$. Thus, in the $\constr=0$ case, still more general than the $\U(N)$-sector, the LQG effective corrections to the Friedmann equation produced by the full model coincide with the corrections obtained in the $\U(N)$ reduced sector.

The classical limit for the reduced $\U(N)$ sector of the two-vertex model is given by the small-twist regime \cite{Cendal:2024uzu,Garay:2025cis}. The natural extension to the present full general case is provided by the global twist $\phi$, which depends on all the different twists, and coincides with that in the $\U(N)$ reduced sector. In fact, $\phi$ relates to $\dot{a}/a$ through equation \eqref{eq:dot_a}, thus providing a measure of the global expansion:  small $\phi$ coincides with  a small expansion rate  $\dot{a}/a$.
We also expect the classical limit of the model to be represented by the regime
$a\gg\amin$ in addition to 
small global twist $\phi$.
Taking these limits in \eqref{eq:dot_a}
and $H=0$,
the generalized Friedmann equation~\eqref{eq:master} becomes
\begin{equation}
    \left(\frac{\dot{a}}{{\mathcal{N}}a}\right)^{2} = -\frac{\kappa}{a^{2}}+ \gamma {f},\label{eq:GR}
  \end{equation}
  which is equivalent to the Friedmann equation
    in FLRW with spatial curvature $\kappa$
    and energy density $\varrho$ equal to $f$ up to a constant factor
    (see \cite{Cendal:2024uzu,Garay:2025cis}).
In particular, if $f(a)$ is constant, we recover the cosmological constant,
while for an equation of state $p=w\varrho$,
using the continuity equation to obtain the pressure $p$,
  we obtain $f\propto a^{-3(1+w)}$. 

As expected,
in this classical limit the terms involving the Barbero-Immirzi parameter $\bi$ disappear and one recovers the pure Friedmann equation. 
This confirms that the terms involving
$\bi$
are indeed to be associated with LQG effective corrections
that are responsible for the Big-Bounce behavior
of the system.

\paragraph{Big Bounce.---} Equation~\eqref{eq:master} shows that the classical trajectory bounces, i.e. $\dot a=0$, occur whenever $\Psi$ reaches one of the two roots of its right-hand side,
\begin{equation}
\Psi_\pm=\frac{1\pm\sqrt{1-\bi^4\constr^2\mu^{2}}}{2\bi^2}.
\end{equation}
The bound $a\geq\amin$ \eqref{eq_minimalarea}, guarantees
$0\leq\bi^4\constr^2\mu^2\leq1$. Thus, $\Psi_\pm$ are real and
\mbox{$0\leq\bi^2\Psi_-\leq1/2\leq\bi^2\Psi_+\leq1$}. For $a\gg\amin$ one has
$\mu\sim0$ and hence
$\Psi_-\sim 0$, $\bi^2\Psi_+\sim 1$, while for $a=\amin$,
$\mu=1/(\bi^2 \constr)$ and the two branches merge, $\bi^2 \Psi_\pm=1/2$.
Moreover, if $\constr =0$ 
we have $\Psi_-= 0$ and $\bi^2 \Psi_+= 1$ for $a>\amin$.
All this means that the anisotropy
correction encoded in the last term of \eqref{eq:master} is bounded,
that its effect at big $a$ is equivalent to
  the smoothing out of the anisotropies of the graph, and that it
only becomes appreciable when $a$ approaches $\amin$.
Thus, the LQG effective corrections to the generalized Friedmann equation
\eqref{eq:master} produced by the full two-vertex model coincide
with those obtained in the $\U(N)$-reduced sector of
\cite{Cendal:2024uzu,Garay:2025cis} away from the Big-Bounce.

\begin{figure}
\centering
\begin{overpic}[width=0.9\linewidth]{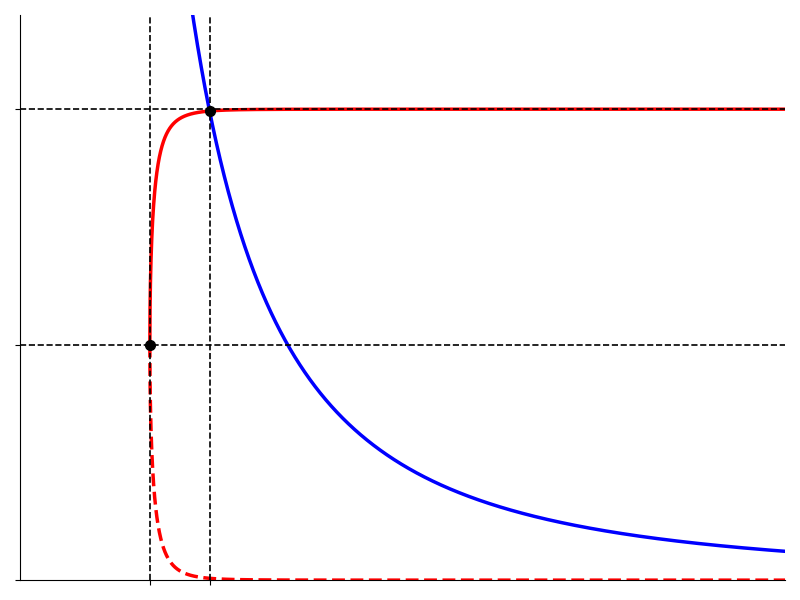}
\put(15,-1){$\amin$}
\put(25,-1){$a_\bb$}
\put(95,-1){$a$}
\put(35,65){\color{red}$\Psi_+$}
\put(40,39){\color{blue}$\Psi$}
\put(35,6){\color{red}$\Psi_-$}
\put(-2,1){$0$}
\put(-6,30.5){$1/2$}
\put(-2,60){$1$}
\end{overpic}
\caption{$\Psi(a)$ (blue) together with the Big-Bounce branches $\Psi_{+}(a)$ (solid red) and $\Psi_{-}(a)$ (dashed red), for the illustrative choice  $\Sigma=\bi=\constr=1$, $\Psi_{0}=3$, $\alpha=2$. The branches merge at $a=\amin$ with $\Psi_{\pm}=1/2$, and separate towards $1$ and $0$, respectively, as $a\to\infty$. The curve $\Psi(a)$ crosses $\Psi_{+}(a)$ at the  Big Bounce scale factor $a=a_\bb$, and remains above $\Psi_{-}(a)$.}
\label{fig:psi-pm}
\end{figure}
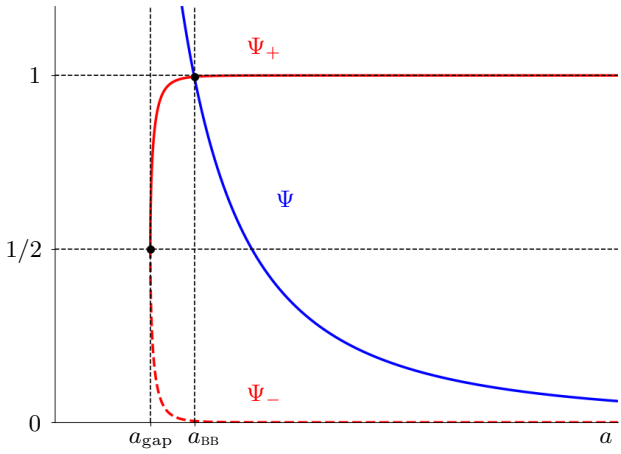

For illustrative purposes 
we consider
$\kappa=0$ and \mbox{$p=w\varrho$} with
$-1/3<w<1$, so $\Psi=\Psi_0\, a^{-\alpha}$ with \mbox{$0<\alpha<4$}.
It can be shown that if $\Psi$ crosses $\Psi_+$, so that there is a Big Bounce, then $\Psi$ never crosses $\Psi_-$ and there is no other bounce.
In Fig.  \ref{fig:psi-pm} we present a numerical example
that shows how $\Psi$ crosses $\Psi_+$ at a value $a_\bb$ that is necessarily larger than
$\amin$.
Thus $a_\bb$ is the Big Bounce, the actual turning point of the scale
factor. 

\paragraph{Conclusion.---}

We have resolved the global behavior of the two-vertex model in LQG, including
the dynamics of non-symmetric sectors previously accessible only via numerical simulations.
By mapping the spinorial microstate space onto a reduced phase space governed by a
Poisson subalgebra, the system's evolution becomes analytically tractable.
This global formulation yields two fundamental geometric features: a lower area bound for the graph acting as a kinematical Big-Bounce threshold, and a generalized Friedmann equation incorporating graph anisotropies and Barbero-Immirzi corrections.
  
Beyond the two-vertex model, this global variable formulation provides a systematic coarse-graining tool for quantum geometry. By compressing vast numbers of microscopic degrees of freedom into physical Casimirs, this framework paves the way for calculating statistical entropy and taking continuum limits on multi-vertex graphs.

\paragraph{Acknowledgments.---}
Financial support provided by the 
Spanish Government grants PID2023-149018NB-C44 (funded by MICIU/AEI/10.13039/501100011033  and ``ERDF/EU A way of making Europe'') 
and PID2021-123226NB-I00 (funded by MCIN/AEI/10.13039/501100011033  and by ``ERDF/EU A way of making Europe''), 
the Basque Government grant IT1977-26. 
Additionally, S. R. acknowledges financial support from MIU (Ministerio de Universidades) fellowship FPU23/01491.

\bibliography{references.bib}

\end{document}